\documentstyle{article}

\begin{document}

\title{A Hydrodynamic Approach \\ to Superconductivity}
\author{Girish S. Setlur \\ Harish Chandra Research Institute
 \\ Chhatnag Road, Jhunsi, Allahabad, India 211 019. }
\maketitle

\begin{abstract}
 Recently Tsai et.al. (cond-mat/0406174) have used the renormalization group
 approach to study strong coupling superconductors without assuming a
 broken symmetry phase. We use the hydrodynamic formulation to 
 study the same problem with the same intention.
 We recast the electron-phonon + electron-electron
 problem in the hydrodynamic language and compute the
 one-particle electron Green function at finite temperature.
 From this we extract the dynamical density of states at finite temperature
 and look for sign of a gap.
\end{abstract}

\section{ Introduction }

\vspace{0.2in}

 Motivated by a series of recent experiments\cite{Exp1}, \cite{Exp2},\cite{Exp3}, \cite{Exp4}, \cite{Exp5}
 Tsai et.al\cite{Tsai} have studied the question of reproducing the results of
 Eliashberg's strong coupling theory of superconductivity without making an
 assumption of a broken symmetry phase. The need for this stems from fact that
 in real systems, there may be many sources of instability and there is no
 general guide (other than through experiments) to determine which of these
 instabilities dominate. Thus from a theoretical standpoint, as Tsai
 et.al. \cite{Tsai} argue, it is desirable to have a theory that is unbiased in the sense that it makes no assumption
 about the system being in a given phase. Rather all sources instability are
 treated on an equal footing and a sufficiently powerful theory should be able to
 pick out which of these phases dominate in what regions of the coupling
 constant/temperature plane. The RG approach of Shankar\cite{shankar} as
 applied to the problem of strongly coupled superconductors by Tsai
 et.al. \cite{Tsai} is one such. Here we present an alternative to this
 interesting and important work, namely the
 hydrodynamic approach. In the hydrodynamic approach, we simply recast the
 electron-phonon + electron-electron problem in terms of the electron's hydrodynamic variables namely the
 density fluctuations and the conjugate namely the velocity potential.
 The crucial fermionic nature of the electron is captured by a nontrivial
 phase functional that enters into the framework in the lagrangian
 formulation. The hamiltonian is identical to the Dashen-Sharp formula in terms of currents and densites.

  Unlike the RG approach\footnote{which the author claims no expertise of}, which requires dividing
 the Fermi surface up into patches, we do not have to pay conscious
 attention to the Fermi surface. The properties of the free Fermi theory are
 automatically encoded in the phase functional. However our approach is by no
 means exact. The phase functional that encodes Fermi statistics has to be
 determined by expanding in powers of density fluctuations. We retain only the
 leading linear term, thereby implying that three-body density correlations are
 implicitly ignored. This means we have to work in the high density limit,
 where the Migdal's theorem applies. This is exactly the regime in which Tsai
 et. al. operate. However, our approach can be generalised in principle to
 regimes of lower density as well by including three-body terms and so on
 (but the calculations are obviously harder
 to carry out in practice). Thus the purpose of the present article is to
 highlight the usefulness of the hydrodynamic approach by comparing favorably
 with the results of well-established methods authored by famous
 physicists\footnote{sardonic tone unintentional}.

\section{The Hydrodynamic Action}

In an earlier preprint\cite{presup}, we recall that we had written down the action for
electrons coupled to phonons in the hydrodynamic language. We rederive that
here for the sake of completeness. We assume following Tsai et.al.\cite{Tsai}
that the electron-electron interaction is short-ranged as is the
electron-phonon interaction. In their work, the bare electron-electron coupling
constant is denoted by $ u_{0} $ and the bare electron-phonon coupling $ g $.
We try to follow the notation of Tsai et.al. as closely possible. The full
action may be written down as follows.
\[
S = \int^{ -i\beta}_{0}dt \mbox{   }\sum_{ \sigma }
 \int d^3 x \mbox{         }
 \mbox{       }\psi^{*}({\bf{x}}\sigma,t)
\mbox{     }\left(i \partial_{t} + \frac{ \nabla^2 }{ 2m_{e} }\right)
\psi({\bf{x}}\sigma,t)
+ \int^{ -i\beta}_{0}dt \mbox{   }\sum^{'}_{ {\bf{q}} } \phi^{*}_{ {\bf{q}} }(t) \left( i
\partial_{t} - w_{ {\bf{q}} } \right) \phi_{ {\bf{q}} }(t) 
\]
\begin{equation}
+  \frac{ g }{ \sqrt{V} }\mbox{     }\int^{ -i\beta}_{0}dt \mbox{   }\sum^{'}_{ {\bf{q}}
}\mbox{      }\rho_{ {\bf{q}} }(t)
\left( \phi_{ {\bf{q}} }(t) + \phi^{*}_{ -{\bf{q}} }(t) \right)
- \frac{ u_{0} }{2V} \int^{ -i\beta}_{0}dt \mbox{   }\sum_{ {\bf{q}}
}\mbox{      }\rho_{ {\bf{q}} }(t)
\rho_{ -{\bf{q}} }(t)
\end{equation}
The prime over the summation $ {\bf{q}} $ indicates that we restrict the values
of $ |{\bf{q}}| < \Lambda_{D} $ the Debye cutoff.
The free part may be recast in the hydrodynamic language as follows.
\begin{equation}
S_{free} =  \int\mbox{        }
\left( \rho \mbox{   } \partial_{t} \Pi - \rho \mbox{   } \partial_{t} \Phi_{e}
 - \frac{ \frac{ (\nabla \rho)^2 }{ 4 \rho } + \rho ( \nabla \Pi)^2 }{ 2m_{e}
 } \right)
\end{equation}
Here the phase function $ \Phi_{e}([\rho];{\bf{x}}\sigma) $  encodes Fermi
statistics. Without it, we would be describing bosons rather than
fermions. Also the Fermi current is given by the hydrodynamic expression which
implicitly defines $ \Pi $ namely, $ {\bf{J}} = - \rho \nabla \Pi $. The phase
functional $ \Phi_{e} $  has to be fixed by making contact with the
correlation functions of the free Fermi theory. This has been done in an
earlier preprint and we may write down the leading term in the expansion in
powers of the density fluctuations. 
\begin{equation}
\Phi_{e}([\rho];{\bf{x}}\sigma) = \sum_{ {\bf{q}} n } e^{ i {\bf{q.x}} }
 e^{ z_{n} t } \mbox{         }C({\bf{q}},n\sigma) \mbox{         } \rho_{ -{\bf{q}}\sigma, -n }
\end{equation}

\begin{equation}
\beta\mbox{     }z_{n}\mbox{         }C({\bf{q}},n\sigma) = \frac{1}{ 2 \left<
  \rho_{ {\bf{q}}\sigma, n } \rho_{ -{\bf{q}}\sigma, -n } \right>_{0} }
 - \frac{ \beta z^2_{n} }{ 2 N^{0} \epsilon_{ {\bf{q}} } }
- \frac{ \beta \epsilon_{ {\bf{q}} } }{ 2 N^{0} }
\end{equation}
Here  $ N^{0} $ is the total number of electrons and,
\begin{equation}
\left< \rho_{ {\bf{q}}\sigma }(t) \rho_{ -{\bf{q}}\sigma }(t^{'}) \right>_{0}
 = \sum_{n} e^{ z_{n}(t-t^{'}) } \mbox{       }
 \left< \rho_{ {\bf{q}}\sigma, n } \rho_{ -{\bf{q}}\sigma, -n } \right>_{0}
\end{equation}
is the density correlation function of the free Fermi theory.
The slow part of the field variable is given by,
\begin{equation}
\psi_{slow}({\bf{x}}\sigma,t)  = e^{ 2 i \sum_{ {\bf{q}} n }e^{ i {\bf{q}}.{\bf{x}} }
 e^{ z_{n} t } C({\bf{q}},n,\sigma) \rho_{ -{\bf{q}}\sigma, -n } }
 e^{ -i \sum_{ {\bf{q}}, n } e^{ i {\bf{q.x}} } e^{ z_{n} t } X_{
 {\bf{q}}\sigma, n } }
\end{equation}
Also $ z_{n} = \frac{ 2 \pi n }{ \beta } $ is the bosonic Matsubara
frequency. Furthermore, the mysterious factor of two in the exponential is to ensure that the
exponents in the one-dimenional case come out right. By expanding in powers of
the density fluctuations and retaining the leading terms we may write the full
action as follows.
\[
S = \sum_{ {\bf{q}}n, \sigma } (-i \beta z_{n}) \mbox{     }\rho_{ {\bf{q}}, n,\sigma }
 X_{ {\bf{q}}, n, \sigma } 
 + \sum_{ {\bf{q}}n, \sigma } (i \beta z_{n}) \mbox{     }
C({\bf{q}},n,\sigma) \mbox{     }
\rho_{ {\bf{q}}, n, \sigma } \mbox{     }
\rho_{ -{\bf{q}}, -n,\sigma }  
\]
\[
 + \frac{ i \beta N^{0} }{2} \sum_{ {\bf{q}}n, \sigma }\epsilon_{ {\bf{q}}
 }\mbox{     }X_{ {\bf{q}}, n, \sigma } X_{ -{\bf{q}}, -n, \sigma } 
+ (-i \beta) \mbox{   }\sum^{'}_{ {\bf{q}}n } \phi^{*}_{ {\bf{q}}, n }
 \left( iz_{n} - w_{ {\bf{q}} } \right) \phi_{ {\bf{q}}, n } 
\]
\begin{equation}
-  \frac{ i \beta g }{ \sqrt{V} }\mbox{   }\sum^{'}_{ {\bf{q}} n }\mbox{      }\rho_{ {\bf{q}}, n }
\left( \phi_{ {\bf{q}}, n } + \phi^{*}_{ -{\bf{q}}, -n } \right)
+ \frac{ i \beta u_{0} }{2V}  \mbox{   }\sum_{ {\bf{q}} n }\mbox{      }\rho_{ {\bf{q}}, n }
\rho_{ -{\bf{q}}, -n }
\end{equation}
Here $ \rho_{ {\bf{q}}, n } = \rho_{ {\bf{q}} \uparrow, n } + \rho_{ {\bf{q}} \downarrow, n } $.
Following Tsai et.al. we choose the phonons to be Einstein-like namely constant
dispersion $ w_{ {\bf{q}} } = w_{E} $. Since the action is purely quadratic in
the phonons, we may integrate it out and write down an effective action for the electrons.
\[
S_{eff} = \sum_{ {\bf{q}}n, \sigma } (-i \beta z_{n}) \mbox{     }\rho_{ {\bf{q}}, n,\sigma }
 X_{ {\bf{q}}, n, \sigma } 
 + \sum_{ {\bf{q}}n, \sigma } (i \beta z_{n}) \mbox{     }
C({\bf{q}},n,\sigma) \mbox{     }
\rho_{ {\bf{q}}, n, \sigma } \mbox{     }
\rho_{ -{\bf{q}}, -n,\sigma }  
\]
\begin{equation}
 + \frac{ i \beta N^{0} }{2} \sum_{ {\bf{q}}n, \sigma }\epsilon_{ {\bf{q}}
 }\mbox{     }X_{ {\bf{q}}, n, \sigma } X_{ -{\bf{q}}, -n, \sigma } 
+ \frac{ i \beta }{2V}  \mbox{   }\sum^{'}_{ {\bf{q}} n }
\left( u_{0} - \frac{  2g^2 w_{E} }{ w_{E}^2 + z_{n}^2 } \right)\mbox{      }\rho_{ {\bf{q}}, n }
\rho_{ -{\bf{q}}, -n }
\end{equation}
Now we wish to compute the propagator. More specifically, we wish to simply
compute the momentum distribution and from that extract the quasiparticle
residue at finite temperature.  Tsai et. al. use this procedure to ascertain
the transition temperature for the metal-superconductor transition.
However the crucial point is that the vanishing of the quasiparticle residue
is a necessary but not sufficient condition for the system to be a
superconductor. It could also be an insulator. Thus either one has to
demonstrate the divergence of the d.c. conductivity or more illuminatingly
 demonstrate phase coherence. In other words the nonvanishing nature of Yang's 
 offdiagonal long-range order correlation function. However since we know from
 prior experience, that this is a superconducting transition, we shall content
 ourselves in just computing the momentum distribution. 
 After integrating out the phonons we may write,
\[
S_{eff} = \sum_{ {\bf{q}}n } (-i \beta z_{n}) \mbox{     }\rho_{ {\bf{q}}, n,\uparrow }
 X_{ {\bf{q}}, n, \uparrow } 
 +  \sum_{ {\bf{q}}n } (-i \beta z_{n}) \mbox{     }\rho_{ {\bf{q}}, n,\downarrow }
 X_{ {\bf{q}}, n, \downarrow } 
\]
\[
 + \sum_{ {\bf{q}}n, \sigma } (i \beta z_{n}) \mbox{     }
C({\bf{q}},n,\sigma) \mbox{     }
\rho_{ {\bf{q}}, n, \uparrow } \mbox{     }
\rho_{ -{\bf{q}}, -n,\uparrow }
  + \sum_{ {\bf{q}}n, \sigma } (i \beta z_{n}) \mbox{     }
C({\bf{q}},n,\sigma) \mbox{     }
\rho_{ {\bf{q}}, n, \downarrow } \mbox{     }
\rho_{ -{\bf{q}}, -n,\downarrow }
\]
\[
 + \frac{ i \beta N^{0} }{2} \sum_{ {\bf{q}}n }\epsilon_{ {\bf{q}}
 }\mbox{     }X_{ {\bf{q}}, n, \uparrow } X_{ -{\bf{q}}, -n, \uparrow } 
 + \frac{ i \beta N^{0} }{2} \sum_{ {\bf{q}}n }\epsilon_{ {\bf{q}}
 }\mbox{     }X_{ {\bf{q}}, n, \downarrow } X_{ -{\bf{q}}, -n, \downarrow } 
\]
\begin{equation}
+ \frac{ i \beta }{2V}  \mbox{   }\sum^{'}_{ {\bf{q}} n }
\left( u_{0} - \frac{  2g^2 w_{E} }{ w_{E}^2 + z_{n}^2 } \right)\mbox{      }
(\rho_{ {\bf{q}} \uparrow, n }\rho_{ -{\bf{q}}\uparrow, -n } 
+ 2 \rho_{ {\bf{q}} \downarrow, n }\rho_{ -{\bf{q}}\uparrow, -n }
+ \rho_{ {\bf{q}} \downarrow, n }\rho_{ -{\bf{q}} \downarrow, -n })
\end{equation}
The full propagator may be written as follows.
\[
\left< T \mbox{   }\psi_{slow}({\bf{x}}\uparrow,t)\psi^{\dagger}_{slow}({\bf{x}}^{'}\uparrow,t^{'})\right> 
  = \left< e^{ i \sum_{ {\bf{q}} n }\left( e^{ i {\bf{q}}.{\bf{x}} }
 e^{ z_{n} t }
 -  e^{ i {\bf{q}}.{\bf{x}}^{'} }
 e^{ z_{n} t^{'} } \right)
 \left( 2 C({\bf{q}},n,\uparrow) \rho_{ -{\bf{q}}\uparrow, -n }
 - X_{ {\bf{q}}\uparrow, n } \right) } \right>
\]
\begin{equation}
  = e^{ -\frac{1}{2} \sum_{ {\bf{q}} n }\left( 2 -  e^{ i {\bf{q}}.({\bf{x}}-{\bf{x}}^{'}) }
 e^{ z_{n} (t-t^{'}) }
 -  e^{ i {\bf{q}}.({\bf{x}}^{'}-{\bf{x}}) }
 e^{ z_{n} (t^{'}-t) } \right)E({\bf{q}},n) }
\end{equation}
\begin{equation}
E({\bf{q}},n) = \left( 4 C({\bf{q}},n,\uparrow) \mbox{     }
 C(-{\bf{q}},-n,\uparrow)
\left<  \rho_{
 {\bf{q}}\uparrow, n }
 \rho_{ -{\bf{q}}\uparrow, -n } \right>
 - 4 C(-{\bf{q}},-n,\uparrow) \left<  X_{ {\bf{q}}\uparrow, n }\rho_{
 {\bf{q}}\uparrow, n } \right>
 + \left< X_{ {\bf{q}} \uparrow, n }
 X_{ -{\bf{q}} \uparrow, -n } \right> \right) 
\end{equation}
Now we make use of the trick outlined earlier namely multiply and divide by
the free propagator, in the denominator use the the bosonized version and in
the numerator use the one obtained from elementary considerations. Thus we may write
 (see appendices for definitions notation e.t.c.),
\begin{equation}
\frac{ \left< T \mbox{
    }\psi({\bf{x}}\uparrow,t)\psi^{\dagger}({\bf{x}}^{'}\uparrow,t^{'})\right>
    }{  \left< T \mbox{
    }\psi({\bf{x}}\uparrow,t)\psi^{\dagger}({\bf{x}}^{'}\uparrow,t^{'})\right>_{0} }
    = e^{ -\frac{1}{2} \sum_{ {\bf{q}} n }\left( 2 -  e^{ i {\bf{q}}.({\bf{x}}-{\bf{x}}^{'}) }
 e^{ z_{n} (t-t^{'}) }
 -  e^{ i {\bf{q}}.({\bf{x}}^{'}-{\bf{x}}) }
 e^{ z_{n} (t^{'}-t) } \right) E_{diff}({\bf{q}},n) }
\end{equation}
 Thus we have an asymptotically exact formula for the propagator provided we
 evaluate $ E $. To this end we first integrate out the down spin variables and write,
\begin{equation}
S_{eff, \uparrow} = \sum_{ {\bf{q}}n } (-i \beta z_{n}) \mbox{     }\rho_{ {\bf{q}}, n,\uparrow }
 X_{ {\bf{q}}, n, \uparrow } 
 + \frac{ i \beta N^{0} }{2} \sum_{ {\bf{q}}n }
\epsilon_{ {\bf{q}} }\mbox{ }X_{ {\bf{q}}, n, \uparrow } X_{ -{\bf{q}}, -n, \uparrow } 
+ \sum^{'}_{ {\bf{q}}n } G_{ \uparrow }({\bf{q}},n)
\rho_{ {\bf{q}}\uparrow, n }\mbox{      }\rho_{ -{\bf{q}}\uparrow, -n } 
\end{equation}
where,
\begin{equation}
G({\bf{q}},n) = \frac{   i \beta z_{n}^2 }{ 2 N^{0}\epsilon_{ {\bf{q}} } } 
+ (i \beta z_{n}) \mbox{     }
C({\bf{q}},n,\downarrow) \mbox{     }
 + \frac{ i \beta }{2V}  \mbox{   }
\left( u_{0} - \frac{  2g^2 w_{E} }{ w_{E}^2 + z_{n}^2 } \right) 
\end{equation}

\begin{equation}
G_{ \uparrow }({\bf{q}},n) =  (i \beta z_{n}) \mbox{     }
C({\bf{q}},n,\uparrow) + \frac{ i \beta }{2V}\left( u_{0} - \frac{  2g^2 w_{E}
}{ w_{E}^2 + z_{n}^2 } \right)\mbox{      }
+  \frac{  \beta^2 }{V^2}
\frac{ \left( u_{0} - \frac{  2g^2 w_{E} }{ w_{E}^2 + z_{n}^2 } \right)^2 }{ 4G(-{\bf{q}},-n) }
\end{equation}
We may now read off the correlation functions. 
\begin{equation}
\left< \rho_{ {\bf{q}}, n,\uparrow } \rho_{ -{\bf{q}}, -n \uparrow } \right>
 = \frac{1}{ \frac{\beta z^2_{n} }{ N^{0} \epsilon_{ {\bf{q}} } }
- 2 i G_{ \uparrow }({\bf{q}},n) }
\label{RHORHO}
\end{equation}

\begin{equation}
\left< X_{ {\bf{q}}, n,\uparrow } X_{ -{\bf{q}}, -n \uparrow } \right> =
\frac{1}{ \beta N^{0}\epsilon_{ {\bf{q}} }   
 + i  \frac{ (\beta z_{n})^2 }{ 2 G_{ \uparrow }(-{\bf{q}},-n) } }
\label{XX}
\end{equation}

\begin{equation}
\left< \rho_{ {\bf{q}}\uparrow,n } X_{  {\bf{q}}\uparrow,n } \right> = \frac{
  (\beta z_{n}) }{  2 i\epsilon_{ {\bf{q}} }
  \beta N^{0}\mbox{   } G_{ \uparrow }({\bf{q}},n)
 - (\beta z_{n})^2 }  
\end{equation}

\begin{equation}
\left< \rho_{ {\bf{q}}, n,\uparrow } \rho_{ -{\bf{q}}, -n \uparrow } \right>_{0}
 = \frac{1}{ \frac{\beta z^2_{n} }{ N^{0} \epsilon_{ {\bf{q}} } }
+ (2 \beta z_{n}) \mbox{     }
C({\bf{q}},n,\uparrow) }
\end{equation}

\begin{equation}
\left< X_{ {\bf{q}}, n,\uparrow } X_{ -{\bf{q}}, -n \uparrow } \right>_{0} =
\frac{1}{ \beta N^{0}\epsilon_{ {\bf{q}} }   
 +  \frac{ (\beta z_{n}) }{ 2 \mbox{     }
C({\bf{q}},n,\uparrow) } }
\end{equation}

\begin{equation}
\left< \rho_{ {\bf{q}}\uparrow,n } X_{  {\bf{q}}\uparrow,n } \right>_{0} =
  \frac{ -1 }{ 2 \epsilon_{ {\bf{q}} }
  \beta N^{0} \mbox{     }
C({\bf{q}},n,\uparrow) + \beta z_{n} }  
\end{equation}
We use the current algebra constraint to determine 
the approximate long-wavelength nature of the correlation functions of the
free theory. This means the unknown $ C $  has to be fixed so that 
 the current-current correlation are related in the usual manner to
 density-density correlation functions.
\begin{equation}
{\bf{j}}_{ {\bf{q}}\uparrow, n} = X_{ -{\bf{q}}\uparrow, -n }(i  \frac{ N^{0} }{2}  {\bf{q}}) 
\end{equation}
\begin{equation}
\left< {\bf{j}}_{ {\bf{q}}\uparrow, n} \cdot {\bf{j}}_{ -{\bf{q}}\uparrow, -n} \right>
 = \frac{ (N^{0})^2 {\bf{q}}^2 }{4} \left< X_{ {\bf{q}}\uparrow, n} X_{ -{\bf{q}}\uparrow, -n} \right>
 = k_{F}^2 \left< \rho_{ {\bf{q}}\uparrow, n} \rho_{ -{\bf{q}}\uparrow, -n} \right>
\end{equation}
This means,
\begin{equation}
 \frac{k^2_{F}}{ \frac{\beta z^2_{n} }{ N^{0} \epsilon_{ {\bf{q}} } }
+ (2 \beta z_{n}) \mbox{     }
C({\bf{q}},n,\uparrow) }
 = \frac{(N^{0})^2 {\bf{q}}^2}{ 4 \beta N^{0}\epsilon_{ {\bf{q}} }   
 +  \frac{ (2 \beta z_{n}) }{ C({\bf{q}},n,\uparrow) } }
\end{equation}
Thus we may deduce,
\begin{equation}
C({\bf{q}},n,\uparrow) \approx \frac{ 2 k^2_{F} }{ z_{n} N^{0} (2m)  }
\end{equation}
This in turn means,
\begin{equation}
\left< \rho_{ {\bf{q}}, n,\uparrow } \rho_{ -{\bf{q}}, -n \uparrow } \right>_{0}
 =\frac{1}{ \beta } \frac{ N^{0} \epsilon_{ {\bf{q}} } }{ z^2_{n} + v^2_{F} q^2 }
\end{equation}
Just to verify that this is sensible, we compute the static density-density
correlation at zero temperature which we know is $ (N^{0}/2) q/(2k_{F}) $.
\[
\sum_{n}\left< \rho_{ {\bf{q}}, n,\uparrow } \rho_{ -{\bf{q}}, -n \uparrow } \right>_{0}
 = \int^{ \infty} _{ - \infty }\frac{dn}{ 2m \beta }
 \frac{ N^{0} q^2 }{ z^2_{n} + v^2_{F} q^2 }
\]
\begin{equation}
 = \frac{ N^{0} }{ 2 } \frac{q}{ 2 k_{F} } 
\end{equation}
as required. It is important to point out that there is no such simple
connection between density-density and curent-current correlation functions
for interacting systems, since the four-point functions are not obligated 
 to resemble the noninteracting values that enable the correspondence.
Therefore, in particular one should not look to similarly relate
 Eq.(~\ref{RHORHO}) and Eq.(~\ref{XX}).
Finally, we wish to ascertain that the last correlation function is consistent
 with current algebra.
We know that,
\begin{equation}
\left< \rho_{ {\bf{q}}\uparrow,n } X_{  {\bf{q}}\uparrow,n } \right>_{0} =
\frac{ -z_{n} }{ \beta \mbox{   } v^2_{F}q^2  + \beta z^2_{n} }  
\end{equation}
In other words,
\begin{equation}
- \frac{2}{ i q^2 N^{0} }
{\bf{q}} \cdot  \left< \rho_{ {\bf{q}}\uparrow,n } {\bf{j}}_{ -{\bf{q}}\uparrow, -n} \right>_{0} =
\frac{ -z_{n} }{ \beta \mbox{   } v^2_{F}q^2  + \beta z^2_{n} }  
\end{equation}

\begin{equation}
\left< \rho_{ {\bf{q}}\uparrow } {\bf{q}} \cdot {\bf{j}}_{ -{\bf{q}}\uparrow }
\right>_{0} = \sum_{ {\bf{k}} } {\bf{k.q}} \mbox{         }n_{F}({\bf{k}} + {\bf{q}}/2 ) ( 1 -
n_{F}({\bf{k}}-{\bf{q}}/2) )
 = \sum_{ {\bf{k}} } {\bf{k.q}} \mbox{         }n_{F}({\bf{k}} + {\bf{q}}/2 )
  = -\frac{ N^{0}q^2 }{4}  
\end{equation}
Thus we must have,
\begin{equation}
(- \frac{2}{ i q^2 N^{0} })
( -\frac{ N^{0}q^2 }{4}  )
= \frac{ v_{F}q }{ 2 i v_{F}q }
\end{equation}
an identity. Now we may proceed to evaluate the full propagator.

\section{ Dynamical Density of States }

Here we compute the full propagator (see appendices for more details and
definitions of the various terms). From there we extract the dynamical
density of states. An examination of this should tell us whether or not a gap is persent.
The one-particle spectral function is given by,
\begin{equation}
2 \pi\mbox{   }A({\bf{k}},\omega) = \int^{ \infty}_{ -\infty}dt \mbox{
}e^{ i \omega t } \mbox{          }\left( \left< c_{ {\bf{k}} \uparrow }(t)
c^{\dagger}_{  {\bf{k}} \uparrow }(0) \right> 
+ \left< c^{\dagger}_{  {\bf{k}} \uparrow }(0)  c_{ {\bf{k}} \uparrow }(t)
\right> \right)
\end{equation}
We define the dynamical density of states as,
\begin{equation}
D(\omega) = \frac{1}{V}\sum_{ {\bf{k}} }A({\bf{k}},\omega) 
\end{equation}
 Thus the relevant quantities are the unequal-time, equal-space Green functions :
 $ \left< T \mbox{ }\psi({\bf{x}}\uparrow,t)\psi^{\dagger}({\bf{x}}\uparrow,t^{'})\right> $.
 From the appendices we find that we may simplify these in two spatial dimensions,
\begin{equation}
\frac{ \left< \psi({\bf{x}}\uparrow,t)\psi^{\dagger}({\bf{x}}\uparrow,t^{'})\right>
    }{  \left< \psi({\bf{x}}\uparrow,t)\psi^{\dagger}({\bf{x}}\uparrow,t^{'})\right>_{0} }
    = e^{ - F_{ > }(t-t^{'}) }
\end{equation}
\begin{equation}
\frac{ \left< \psi^{\dagger}({\bf{x}}\uparrow,t^{'})\psi({\bf{x}}\uparrow,t)\right>
    }{  \left< \psi^{\dagger}({\bf{x}}\uparrow,t^{'})\psi({\bf{x}}\uparrow,t)\right>_{0} }
    = e^{ - F_{ < }(t-t^{'}) }
\end{equation}

\begin{equation}
F_{>}(t-t^{'}) \approx  
\frac{ ( {\tilde{u}}_{r} - k_{F} ) }{ \pi \rho^{0} \beta }
\mbox{         }   i\Lambda_{D}(t-t^{'})
\end{equation}

\begin{equation}
F_{<}(t-t^{'}) \approx  
\frac{ ( {\tilde{u}}_{r} - k_{F} ) }{ \pi \rho^{0} \beta }
\mbox{         }   i\Lambda_{D}(t^{'}-t)
\end{equation}

\begin{equation}
 {\tilde{u}}_{r} - k_{F} \approx \frac{ m \rho^{0} ( u_{0} - 2 g^2/w_{E} ) }{2k_{F}}
\end{equation}
To first order, we take the zero temperature free Green functions.
\[
\left< \psi({\bf{x}}\uparrow,t)\psi^{\dagger}({\bf{x}}\uparrow,0)\right>_{0}
= \frac{2 m \pi}{(2\pi)^2} \mbox{        }
\frac{ e^{ - i\epsilon_{F} t } }{ i t }
\]

\[
\left< \psi^{\dagger}({\bf{x}}\uparrow,0)
\psi({\bf{x}}\uparrow,t) \right>_{0}
=  \frac{2 m \pi}{(2\pi)^2} 
\frac{ 1 - e^{ - i\epsilon_{F} t } }{ i t }
\]
In this case,
\[
2 \pi D_{0}(\omega) = \int^{ \infty }_{ - \infty } \frac{ dt }{2 \pi i} \mbox{     }e^{ i \omega
  t }\mbox{        } \frac{m}{ t } = m \mbox{        }\theta(\omega)
\]
Thus,
\begin{equation}
 D_{0}(\omega) =  \frac{ m }{2 \pi} \mbox{        }\theta(\omega)
\end{equation}
as required.  In general,
\begin{equation}
\left< \psi({\bf{x}}\uparrow,t)\psi^{\dagger}({\bf{x}}\uparrow,0)\right>
= \frac{2 m \pi}{(2\pi)^2} \mbox{        }
\frac{ e^{ - i\epsilon_{F} t } }{ i t } e^{ -  \frac{ ( {\tilde{u}}_{r} - k_{F} ) }{ \pi \rho^{0} \beta }
\mbox{         }   i\Lambda_{D}\mbox{  }t }
\end{equation}

\begin{equation}
\left< \psi^{\dagger}({\bf{x}}\uparrow,0)
\psi({\bf{x}}\uparrow,t) \right>
=  \frac{2 m \pi}{(2\pi)^2} 
\frac{ 1 - e^{ - i\epsilon_{F} t } }{ i t }
 e^{ \frac{ ( {\tilde{u}}_{r} - k_{F} ) }{ \pi \rho^{0} \beta }
\mbox{         }   i\Lambda_{D}\mbox{   }t }
\end{equation}
Therefore,
\begin{equation}
2 \pi D_{0}(\omega) = \int^{ \infty }_{ - \infty } \frac{ dt }{2 \pi i} \mbox{
  }e^{ i (\omega - \Delta)
  t }\mbox{        } \frac{m}{ t } = \frac{ m }{2 \pi} \mbox{
  }\theta(\omega - \Delta)
\end{equation}
where,
\begin{equation}
\Delta =  k_{B} T \mbox{       }\frac{  m ( 2 g^2/w_{E} - u_{0} )  \Lambda_{D}  }{ 2 \pi k_{F} }
\end{equation}
We may provisionally identify $ \Delta $ with a gap. Thus we find that there
is no superconductivity unless  the (screened) phonon strength exceeds the 
(screened) electron-electron repulsion. Thus we have the (necessary) condition for 
 superconductivity $  g^2 > u_{0} w_{E}/2 $. In particular we may also
 conclude that for $ g^2 = 0 $ and $ u_{0} < 0 $ also we have a gap and hence
 possibly also superconductivity. Strictly speaking we have to compute the
 d.c. conductivity also and show that it diverges in order to be convinced
 that the transition is superconducting (or demonstrate phase coherence ).
 But since we already know this to be
 a superconducting transition we shall content ourselves with doing the bare
 minimum as do Tsai et.al.

\Large

{\bf{ This preprint is Incomplete }}

I seem to be making some errors in the computation of integrals. Perhaps some
more knowledgeable people can help me out. In any case this is meant to show 
members of hiring commitees that I am despertaely trying to do something important.

\normalsize

\section{ Appendix A }

Here we provide some details of the computations of the propagator.
\begin{equation}
G({\bf{q}},n) = \frac{  2 m i \beta z_{n}^2 }{ 2 N^{0} q^2 } 
+ \frac{ 2 i \beta k^2_{F} }{ N^{0} (2m)  }\mbox{     }
 + \frac{ i \beta }{2V}  \mbox{   }
\left( u_{0} - \frac{  2g^2 w_{E} }{ w_{E}^2 + z_{n}^2 } \right) 
\end{equation}

\begin{equation}
G_{ \uparrow }({\bf{q}},n) =  (i \beta z_{n}) \mbox{     }
\frac{ 2 k^2_{F} }{ z_{n} N^{0} (2m)  } + \frac{ i \beta }{2V}\left( u_{0} - \frac{  2g^2 w_{E}
}{ w_{E}^2 + z_{n}^2 } \right)\mbox{      }
+  \frac{  \beta^2 }{V^2}
\frac{ \left( u_{0} - \frac{  2g^2 w_{E} }{ w_{E}^2 + z_{n}^2 } \right)^2 }{ 4G(-{\bf{q}},-n) }
\end{equation}

\begin{equation}
G_{ \uparrow }({\bf{q}},n) = \frac{ ( i \beta ( - 2 g^2 m \rho^{0} w_{E} ( 2
  k_{F}^2 q^2 + m^2 z_{n}^2 ) + ( w_{E}^2 + z_{n}^2 ) ( 2 k_{F}^2 q^2 (m \rho^{0}
  u_{0} + k_{F}^2 ) + m^2 ( m \rho^{0} u_{0} + 2 k_{F}^2 ) z_{n}^2 ) )) }
{ ( m N^{0} ( - 2 g^2 m \rho^{0} q^2 w_{E} + ( w_{E}^2 + z_{n}^2 ) ( q^2 (m \rho^{0}
  u_{0} + 2 k_{F}^2 ) + 2 m^2 z_{n}^2 ))) }
\end{equation}

\begin{equation}
G_{ \uparrow }({\bf{q}},n) = \frac{ i \beta }{m N^{0}} \frac{ - 2 g^2 m \rho^{0} w_{E} ( 2
  v_{F}^2 q^2 + z_{n}^2 ) + ( w_{E}^2 + z_{n}^2 ) ( 2 v_{F}^2 q^2 (m \rho^{0}
  u_{0} + k_{F}^2 ) + ( m \rho^{0} u_{0} + 2 k_{F}^2 ) z_{n}^2 )  }
{ - 2 g^2 \rho^{0} \frac{ q^2 }{m} w_{E} + ( w_{E}^2 + z_{n}^2 ) ( \frac{ q^2 }{m} (\rho^{0}
  u_{0} + \frac{ 2 k_{F}^2 }{m} ) + 2 z_{n}^2 ) }
\end{equation}

\begin{equation}
\left< \rho_{ {\bf{q}}, n,\uparrow } \rho_{ -{\bf{q}}, -n \uparrow } \right>
 = \frac{1}{ \frac{\beta z^2_{n} }{ N^{0} \epsilon_{ {\bf{q}} } }
- 2 i G_{ \uparrow }({\bf{q}},n) }
\end{equation}

\begin{equation}
\left< X_{ {\bf{q}}, n,\uparrow } X_{ -{\bf{q}}, -n \uparrow } \right> =
\frac{1}{ \beta N^{0}\epsilon_{ {\bf{q}} }   
 + i  \frac{ (\beta z_{n})^2 }{ 2 G_{ \uparrow }(-{\bf{q}},-n) } }
\end{equation}

\begin{equation}
\left< \rho_{ {\bf{q}}\uparrow,n } X_{  {\bf{q}}\uparrow,n } \right> =
 \frac{ (\beta z_{n}) }{  2 i\epsilon_{ {\bf{q}} }
  \beta N^{0}\mbox{   } G_{ \uparrow }({\bf{q}},n)
 - (\beta z_{n})^2 }  
\end{equation}

\tiny
\begin{equation}
\left< \rho_{ {\bf{q}}, n,\uparrow } \rho_{ -{\bf{q}}, -n \uparrow } \right>
 = \frac{ N^{0} }{ \beta } 
\left( \frac{ z_{n}^2 }{ \epsilon_{q} } 
 + \frac{ 4 g^2 m N^{0} w_{E} ( 2 k^2_{F} q^2 + m^2 z_{n}^2 )
 - 2 ( w^2_{E} + z_{n}^2 ) ( 2 k^2_{F} q^2 ( m N^{0} u_{0} + k^2_{F} V )
 + m^2 ( m N^{0} u_{0} + 2 k^2_{F} V) z_{n}^2 ) }
{ m ( 2 g^2 m N^{0} q^2 w_{E} -  ( w^2_{E} + z_{n}^2 ) ( q^2 ( m N^{0} u_{0} + 2
 k^2_{F} V ) + 2 m^2 V z_{n}^2 ) ) } \right)^{-1}
\end{equation}

\normalsize
\begin{equation}
\left< \rho_{ {\bf{q}}, n,\uparrow } \rho_{ -{\bf{q}}, -n \uparrow } \right>_{0}
 = \frac{ N^{0} \epsilon_{q} }{ \beta } 
\left( z_{n}^2 + v^2_{F} q^2  \right)^{-1}
\end{equation}

\tiny
\begin{equation}
Ediff({\bf{q}},n) =\hspace{-0.02in} -\frac{\beta(-16m^4n^4\pi^4+8\beta^2k_{F}^2m^2n^2\pi^2q^2+\beta^4k_{F}^4q^4)(4n^2\pi^2u_{0}+\beta^2w_{E}(-2g^2+u_{0}w_{E}))}{\hspace{-0.04in}4n^2\pi^2(4m^2n^2\pi^2+\beta^2 k_{F}^2 q^2)
(16m^2n^4\pi^4V+\beta^4q^2w_{E}(-2g^2mN^{0}+mN^{0}u_{0}w_{E}+k_{F}^2Vw_{E})+4\beta^2n^2\pi^2
(q^2(m N^{0} u_{0}+k_{F}^2 V)+m^2Vw_{E}^2))}
\end{equation}

\normalsize
\begin{equation}
\frac{ \left< T \mbox{
    }\psi({\bf{x}}\uparrow,t)\psi^{\dagger}({\bf{x}}^{'}\uparrow,t^{'})\right>
    }{  \left< T \mbox{
    }\psi({\bf{x}}\uparrow,t)\psi^{\dagger}({\bf{x}}^{'}\uparrow,t^{'})\right>_{0} }
    = e^{ -\frac{1}{2} \sum_{ {\bf{q}} n }\left( 2 -  e^{ i {\bf{q}}.({\bf{x}}-{\bf{x}}^{'}) }
 e^{ z_{n} (t-t^{'}) }
 -  e^{ i {\bf{q}}.({\bf{x}}^{'}-{\bf{x}}) }
 e^{ z_{n} (t^{'}-t) } \right)( E({\bf{q}},n) - E_{0}({\bf{q}},n) )}
\end{equation}
Consider the following identity from complex analysis.
\begin{equation} 
 \oint_{c.p.} \frac{dz}{ e^{ 2 \pi i z } - 1 } f(z) =
\sum_{n = -\infty}^{ \infty } f(n)  + \sum_{ m = all.poles.of.f } \frac{ f_{r}(z_{m}) }{   e^{ 2 \pi i z_{m} } - 1 }
\end{equation}
\begin{equation} 
 f_{r}(z_{m}) = (2 \pi i) \mbox{       }Lt_{ z \rightarrow z_{m} }\mbox{       }(z-z_{m})f(z)
\end{equation}
If $ f $ falls off fast enough,
\begin{equation} 
\sum_{n = -\infty}^{ \infty } f(n) = -\sum_{ m = all.poles.of.f } \frac{ f_{r}(z_{m}) }{   e^{ 2 \pi i z_{m} } - 1 }
\end{equation}
We note that the pathological $ n = 0 $ should be excluded from
consideration. The reasons for this are not entirely clear but doing so
enables the right exponent of the Hubbard model to be recovered\cite{Hubbhydro}. 
When $ m = 1, 2 $ we have,
\begin{equation}
f_{r}(z_{1,2}) =\hspace{-0.02in}
-\frac{(2 \pi i)\beta(-16m^4n_{1}^4\pi^4+8\beta^2k_{F}^2m^2n_{1}^2\pi^2q^2+\beta^4k_{F}^4q^4)
(4n_{1}^2\pi^2u_{0}+\beta^2w_{E}(-2g^2+u_{0}w_{E}))}{\hspace{-0.04in}4n_{1}^2\pi^2(8m^2n_{1}\pi^2)
(16m^2n_{1}^4\pi^4V+\beta^4q^2w_{E}(-2g^2mN^{0}+mN^{0}u_{0}w_{E}+k_{F}^2Vw_{E})+4\beta^2n_{1}^2\pi^2
(q^2(m N^{0} u_{0}+k_{F}^2 V)+m^2Vw_{E}^2))}
\end{equation}

\begin{equation}
z_{1} = i \frac{ \beta k_{F} q }{ 2 m \pi }
\end{equation}
\begin{equation}
z_{2} = - i \frac{ \beta k_{F} q }{ 2 m \pi }
\end{equation}

\begin{equation}
f_{r}(z_{3,4}) =\hspace{-0.02in}
-\frac{(2 \pi i)\beta(-16m^4n_{3}^4\pi^4+8\beta^2k_{F}^2m^2n_{3}^2\pi^2q^2
+\beta^4k_{F}^4q^4)(4n_{3}^2\pi^2u_{0}+\beta^2w_{E}(-2g^2+u_{0}w_{E}))}
{\hspace{-0.04in}4n_{3}^2\pi^2(4m^2n_{3}^2\pi^2+\beta^2 k_{F}^2 q^2)
(64m^2n_{3}^3\pi^4V+8\beta^2n_{3}\pi^2
(q^2(m N^{0} u_{0}+k_{F}^2 V)+m^2Vw_{E}^2))}
\end{equation}

\begin{equation}
z_{3} = i \frac{ \beta w_{E} }{ 2 \pi }
\end{equation}

\begin{equation}
z_{4} = - i \frac{ \beta w_{E} }{ 2 \pi }
\end{equation}

\begin{equation}
f_{r}(z_{5,6}) =\hspace{-0.02in} -\frac{(2 \pi i)
\beta(-16m^4n_{5}^4\pi^4+8\beta^2k_{F}^2m^2n_{5}^2\pi^2q^2+\beta^4k_{F}^4q^4)
(4n_{5}^2\pi^2u_{0}+\beta^2w_{E}(-2g^2+u_{0}w_{E}))}{\hspace{-0.04in}4n_{5}^2\pi^2
(4m^2n_{5}^2\pi^2+\beta^2 k_{F}^2 q^2)
(64m^2n_{5}^3\pi^4V+8\beta^2n_{5}\pi^2
(q^2(m N^{0} u_{0}+k_{F}^2 V)+m^2Vw_{E}^2))}
\end{equation}

\begin{equation}
z_{5} = i\frac{ \beta q }{ 2 \pi m \sqrt{ w_{E} } } 
( - 2 g^2 m \rho^{0} + m \rho^{0} u_{0} w_{E} + k_{F}^2 w_{E} )^{\frac{1}{2}}
 = \frac{ i \beta }{2 \pi }v_{*} q
\end{equation}

\begin{equation}
z_{6} = - i\frac{ \beta q }{ 2 \pi m \sqrt{ w_{E} } } 
( - 2 g^2 m \rho^{0} + m \rho^{0} u_{0} w_{E} + k_{F}^2 w_{E} )^{\frac{1}{2}}
 = - \frac{ \beta i }{ 2 \pi } v_{*} q
\end{equation}

\begin{equation}
f_{r}(z_{1}) = -\frac{ k_{F} }{ N^{0} q }
\end{equation}

\begin{equation}
f_{r}(z_{2}) = \frac{ k_{F} }{ N^{0} q }
\end{equation}

\begin{equation}
f_{r}(z_{3}) = \frac{ g^2 }{ V w_{E}^2 }
\end{equation}

\begin{equation}
f_{r}(z_{4}) = -\frac{ g^2 }{ V w_{E}^2 }
\end{equation}

\begin{equation}
f_{r}(z_{5}) = -\frac{1}{q}  \frac{ \left( - 4 g^4 m^2 (\rho^{0})^2 + 4 g^2 m \rho^{0}
( 2 k_{F}^2 + m \rho^{0} u_{0} ) w_{E} 
 - ( 2 k^4_{F} + 4 k_{F}^2 m \rho^{0} u_{0} + m^2 (\rho^{0})^2 u_{0}^2 ) w_{E}^2
 \right) }{ ( 2 \rho^{0} V \sqrt{ w_{E} } ( - 2 g^2 m \rho^{0} + ( k^2_{F} + m
  \rho^{0} u_{0} ) w_{E} )^{\frac{3}{2}} ) } \equiv \frac{ {\tilde{u}}_{r} }{N^{0}
  q}
\end{equation}

\begin{equation}
f_{r}(z_{6}) = -\frac{1}{q} \frac{ \left( 4 g^4 m^2 (\rho^{0})^2 - 4 g^2 m
  \rho^{0} ( 2 k^2_{F} + m  \rho^{0} u_{0} ) w_{E} + ( 2 k^4_{F} + 4 k^2_{F} m
  \rho^{0} u_{0} + m^2 (\rho^{0})^2 u_{0}^2 ) w_{E}^2 \right) }{ ( 2 \rho^{0} V \sqrt{
  w_{E} } ( - 2 g^2 m \rho^{0} + ( k_{F}^2 + m \rho^{0} u_{0} ) w_{E} )^{\frac{3}{2}} ) } \equiv -\frac{ {\tilde{u}}_{r} }{N^{0} q}
\end{equation}

\begin{equation}
\frac{ \left< \psi^{\dagger}({\bf{x}}^{'}\uparrow,t^{+})\psi({\bf{x}}\uparrow,t)\right>  }
{ \left<\psi^{\dagger}({\bf{x}}^{'}\uparrow,t^{+})
\psi({\bf{x}}\uparrow,t)
\right>_{0} } = e^{ -\sum_{ {\bf{q}} }\left( 1 - cos[ {\bf{q}}.({\bf{x}}-{\bf{x}}^{'})
  ] \right)\sum_{n} f({\bf{q}},n)}
\end{equation}

\begin{equation}
\sum_{n = -\infty}^{ \infty }E_{diff}({\bf{q}},n)  = - \frac{ -\frac{ k_{F} }{ N^{0} q } }{ e^{ -\beta v_{F} q  } - 1 }
 - \frac{ \frac{ k_{F} }{ N^{0} q } }{ e^{ \beta v_{F} q } - 1 }
- \frac{  \frac{ g^2 }{ V w_{E}^2 } }{ e^{ - \beta w_{E} } - 1 }
- \frac{  -\frac{ g^2 }{ V w_{E}^2 } }{ e^{ \beta w_{E} } - 1 }
- \frac{ \frac{ {\tilde{u}}_{r} }{ N^{0} q} }{ e^{ - \beta v_{*} q } - 1 }
- \frac{ -\frac{ {\tilde{u}}_{r} }{ N^{0}q} } { e^{ \beta v_{*} q } - 1 }
\end{equation}
In the noninteracting limit, this sum is zero since $ v_{*} \rightarrow v_{F} $
and $ {\tilde{u}}_{r} \rightarrow k_{F} $ and $ g^2  \rightarrow 0 $.

\section{ Appendix B }

Here we calculate the equal-space unequal-time Green functions.
\begin{equation}
\frac{ \left< T \mbox{
    }\psi({\bf{x}}\uparrow,t)\psi^{\dagger}({\bf{x}}\uparrow,t^{'})\right>
    }{  \left< T \mbox{
    }\psi({\bf{x}}\uparrow,t)\psi^{\dagger}({\bf{x}}\uparrow,t^{'})\right>_{0} }
    = e^{ -\frac{1}{2} \sum_{ {\bf{q}} n }
\left( 2 - e^{ z_{n} (t-t^{'}) } -  e^{ z_{n} (t^{'}-t) } \right)( E({\bf{q}},n) - E_{0}({\bf{q}},n) )}
\end{equation}
Let us first assume $ Im[ t-t^{'} ] < 0 $. This means,
\begin{equation}
\frac{ \left< \psi({\bf{x}}\uparrow,t)\psi^{\dagger}({\bf{x}}\uparrow,t^{'})\right>
    }{  \left< \psi({\bf{x}}\uparrow,t)\psi^{\dagger}({\bf{x}}\uparrow,t^{'})\right>_{0} }
    = e^{ -\sum_{ {\bf{q}} < \Lambda_{D} }F_{>}({\bf{q}},t-t^{'}) }
\end{equation}
where,
\begin{equation}
F_{>}({\bf{q}},t-t^{'}) = \frac{1}{2} \sum_{ n }
\left( 2 - e^{ z_{n} (t-t^{'}) } -  e^{ z_{n} (t^{'}-t) } \right)\mbox{      }E_{diff}({\bf{q}},n) 
\end{equation}

\[
F_{>}({\bf{q}},t-t^{'}) =  \left( - \frac{ -\frac{ k_{F} }{ N^{0} q } }{ e^{ -\beta v_{F} q  } - 1 }
 - \frac{ \frac{ k_{F} }{ N^{0} q } }{ e^{ \beta v_{F} q } - 1 }\right)
\mbox{         }\left(1-e^{ -iv_{F}q (t-t^{'}) }\right)
\]
\[
+\left( - \frac{  \frac{ g^2 }{ V w_{E}^2 } }{ e^{ - \beta w_{E} } - 1 }
- \frac{  -\frac{ g^2 }{ V w_{E}^2 } }{ e^{ \beta w_{E} } - 1 }\right)
\mbox{         }\left(1-e^{ -i w_{E} (t-t^{'}) }\right)
\]
\begin{equation}
+ \left( - \frac{ \frac{ {\tilde{u}}_{r} }{ N^{0} q} }{ e^{ - \beta v_{*} q } - 1 }
- \frac{ -\frac{ {\tilde{u}}_{r} }{ N^{0}q} } { e^{ \beta v_{*} q } - 1 }
\right)\mbox{         }\left(1-e^{ -i v_{*}q (t-t^{'}) }\right)
\end{equation}
Next we assume $ Im[ t-t^{'} ] > 0 $ then,
\begin{equation}
\frac{ \left<\psi^{\dagger}({\bf{x}}\uparrow,t^{'}) \psi({\bf{x}}\uparrow,t)  \right>
    }{  \left< \psi^{\dagger}({\bf{x}}\uparrow,t^{'})\psi({\bf{x}}\uparrow,t) \right>_{0} }
    = e^{ -\sum_{ {\bf{q}}  < \Lambda_{D}  }F_{<}({\bf{q}},t-t^{'}) }
\end{equation}
where,
\begin{equation}
F_{<}({\bf{q}},t-t^{'}) = \frac{1}{2} \sum_{ n }
\left( 2 - e^{ z_{n} (t-t^{'}) } -  e^{ z_{n} (t^{'}-t) } \right)\mbox{      }E_{diff}({\bf{q}},n) 
\end{equation}

\[
F_{<}({\bf{q}},t-t^{'}) =  \left( - \frac{ -\frac{ k_{F} }{ N^{0} q } }{ e^{ -\beta v_{F} q  } - 1 }
 - \frac{ \frac{ k_{F} }{ N^{0} q } }{ e^{ \beta v_{F} q } - 1 }\right)
\mbox{         }\left(1-e^{ iv_{F}q (t-t^{'}) }\right)
\]
\[
+\left( - \frac{  \frac{ g^2 }{ V w_{E}^2 } }{ e^{ - \beta w_{E} } - 1 }
- \frac{  -\frac{ g^2 }{ V w_{E}^2 } }{ e^{ \beta w_{E} } - 1 }\right)
\mbox{         }\left(1-e^{ i w_{E} (t-t^{'}) }\right)
\]
\begin{equation}
+ \left( - \frac{ \frac{ {\tilde{u}}_{r} }{ N^{0} q} }{ e^{ - \beta v_{*} q } - 1 }
- \frac{ -\frac{ {\tilde{u}}_{r} }{ N^{0}q} } { e^{ \beta v_{*} q } - 1 }
\right)\mbox{         }\left(1-e^{ i v_{*}q (t-t^{'}) }\right)
\end{equation}
The momentum integrals cannot be done exactly, hence we make some
assumptions. We assume that the critical temperature lies betwen the Debye
energy and the phonon energy. In other words, $ v_{F}\Lambda_{D} \ll k_{B} T_{c} \ll
 w_{E} $. This means,
\[
F_{>}({\bf{q}},t-t^{'}) \approx \left( - \frac{ 2k_{F}/v_{F} }{ N^{0} \beta q^2 }\right)
\mbox{         }\left(1-e^{ -iv_{F}q (t-t^{'}) }\right)
+ \left( \frac{ 2{\tilde{u}}_{r}/v_{*} }{ N^{0} \beta q^2 }\right)
\mbox{         }\left(1-e^{ -iv_{*}q (t-t^{'}) }\right)
\]
\begin{equation}
+\left( \frac{ g^2 }{ V w_{E}^2 } \right)
\mbox{         }\left(1-e^{ -i w_{E} (t-t^{'}) }\right)
\end{equation}

\[
F_{<}({\bf{q}},t-t^{'}) \approx \left( - \frac{ 2k_{F}/v_{F} }{ N^{0} \beta q^2 }\right)
\mbox{         }\left(1-e^{ iv_{F}q (t-t^{'}) }\right)
+ \left( \frac{ 2{\tilde{u}}_{r}/v_{*} }{ N^{0} \beta q^2 }\right)
\mbox{         }\left(1-e^{ iv_{*}q (t-t^{'}) }\right)
\]
\begin{equation}
+\left( \frac{ g^2 }{ V w_{E}^2 } \right)
\mbox{         }\left(1-e^{ i w_{E} (t-t^{'}) }\right)
\end{equation}
First we focus on two dimensions. We have to make use of the temperature constraint
repaeatedly namely $ v_{F} \Lambda_{D} \ll k_{B} T \ll w_{E} $. Then,
\[
F_{>}(t-t^{'}) \approx \frac{1}{2\pi} 
\left( - \frac{ 2k_{F}/v_{F} }{ \rho^{0} \beta }\right)
\mbox{         }  \int^{ \Lambda_{D} }_{0} dq\mbox{         }
\frac{ 1-e^{ -iv_{F}q (t-t^{'}) } }{q}
+  \frac{1}{2\pi} 
\left( \frac{ 2{\tilde{u}}_{r}/v_{*} }{ \rho^{0} \beta }\right)
\mbox{         }
  \int^{ \Lambda_{D} }_{0} dq\mbox{         }
\frac{ 1 - e^{ -iv_{*}q (t-t^{'}) } }{q}
\]
\begin{equation}
+\left( \frac{ g^2 }{ w_{E}^2 } \right)\frac{ \pi \Lambda_{D}^2 }{(2\pi)^2} 
\end{equation}
 Since $ t-t^{'} \sim 1/(k_{B} T) $ the temperature constraint tells us that
 we may expand in powers of small $ q $.
\begin{equation}
F_{>}(t-t^{'}) \approx  
\frac{ ( {\tilde{u}}_{r} - k_{F} ) }{ \pi \rho^{0} \beta }
\mbox{         }   i\Lambda_{D}(t-t^{'})
+\left( \frac{ g^2 }{ w_{E}^2 } \right)\frac{ \pi \Lambda_{D}^2 }{(2\pi)^2} 
\end{equation}



\end{document}